# STRUCTURING OF COUNTERIONS AROUND DNA DOUBLE HELIX: A MOLECULAR DYNAMICS STUDY

O.O. LIUBYSH[1], A.V. VLASIUK[2], S.M. PEREPELYTSYA[3]

[1]Taras Shevchenko National University of Kyiv,
64, Volodymyrska Str., Kyiv 01033, Ukraine

[2]The Biotechnology Center of the Technische Universitat
Dresden, 47/49 Tatzberg Str., 01307 Dresden, Germany

[3]Bogolyubov Institute for Theoretical Physics, NAS of Ukraine,
14-b Metrolohichna Str., Kyiv, 03680, Ukraine
perepelytsya@bitp.kiev.ua

Abstract

Structuring of DNA counterions around the double helix has been studied by the molecular dynamics method. A DNA dodecamer d(CGCGAATTCGCG) in water solution with the alkali metal counterions $Na^+$, $K^+$, and $Cs^+$ has been simulated. The systems have been considered in the regimes without excess salt and with different salts (0.5 M of NaCl, KCl or CsCl) added. The results have showed that the $Na^+$ counterions interact with the phosphate groups directly from outside of the double helix and via water molecules at the top edge of DNA minor groove. The potassium ions are mostly localized in the grooves of the double helix, and the cesium ions penetrate deeply inside the minor groove being bonded directly to the atoms of nucleic bases. Due to the electrostatic repulsion the chlorine ions tend to be localized at large distances from the DNA polyanion, but some $Cl^-$ anions have been detected near atomic groups of the double helix forming electrically neutral pairs with counterions already condensed on DNA. The DNA sites, where counterions are incorporated, are characterized by local changes of double helix structure. The lifetime of $Na^+$ and $K^+$ in complex with DNA atomic groups is less than 0.5 ns, while in the case of the cesium ions it may reach several nanoseconds. In this time scale, the $Cs^+$ counterions form a structured system of charges in the DNA minor groove that can be considered as ionic lattice.

## 1. Introduction

Under the natural conditions, the DNA double helix is surrounded by water molecules and positively charged ions (counterions). The counterions (usually, $Na^+$, $K^+$, or $Mg^{2+}$) condense on DNA polyanion neutralizing the negatively charged atomic groups of the macromolecule [1-5]. The neutralization of negative charge by counterions is the key point in stability of DNA double helix. The relative disposition of counterions on macromolecule is modulated by double helical structure of DNA. Structuring of counterions around DNA macromolecule is a necessary stage in many mechanisms of conformational transformation of the double helix, such as DNA condensation, formation of microscopic textures on the surface, and others [6-11]. The study of dynamical structure of counterions is important for understanding the molecular level of biological processes where DNA macromolecule is involved.

The localization of counterions on the macromolecule can be determined definitely only in the case of solid samples of DNA [12-14]. In solution the counterions are mobile, therefore they are detected experimentally as the cloud that they form around DNA double helix [15-18]. At the same time, the molecular dynamics method (MD) can provide detailed information about the interaction of counterions with DNA and about the changes in the double helix structure [19-22]. In this regard the MD study of DNA systems is indispensable for understanding the organization of counterions around the double helix.

Starting with the early MD simulations, the counterions are known to interact to the phosphate groups of double helix backbone, the atoms of nucleic bases, and the oxygen atom of deoxyribose [23-26]. The character of counterion interaction essentially depends on the counterion type, concentration, and sequence of nucleic bases [27-35]. The results for alkali metal ions (Li$^+$, Na$^+$, K$^+$, Rb$^+$, and Cs$^+$) show that the sodium ions interact mostly with the phosphate groups [29, 30]. They can be also localized at the top edge of the minor groove and interact with phosphates via water molecules. The potassium ions are usually localized in the groves of the double helix close to the negative base sites [22]. The heavy ions of Rb$^+$ and Cs$^+$ penetrate deeply inside the minor groove and interact directly to the atoms of nucleic bases and the oxygen atoms of deoxyribose [27, 34, 35]. The smallest ions Li$^+$ squeeze through water molecules surrounding the phosphate group and form stable complexes with phosphate groups of the DNA backbone [27]. The counterion organization around DNA macromolecule induce different structure changes that are the key factors in conformational transitions between *A*-, *B*-, and *Z*-forms of the double helix [36-38].

Due to the regularity of DNA structure the counterions should form a regular structure around the double helix. In the previous studies, the counterions tethered to the phosphate groups of DNA backbone have been showed to form a structure resembling an ionic crystal lattice (ion-phosphate lattice) [39-43]. The existence of ion-phosphate lattice has been confirmed by modes of counterion vibrations with respect to the phosphate groups observed in the Cs-DNA low-frequency spectra ( < 200 cm$^{-1}$) [44]. Effects of counterion ordering have been also observed in conductivity experiments of DNA water solutions with KCl salt [45]. To elucidate the lattice arrangement of counterions around the DNA double helix it is necessary to conduct molecular dynamics studies.

The goal of the present research is to study the counterion structuring around the DNA double helix using the molecular dynamics method. To solve this problem, the MD simulations of DNA with Na$^+$, K$^+$, and Cs$^+$ counterions have been made. The description of modelled systems and MD simulation procedure is given in the second section. In the third section, the typical positions of counterions have been determined by analyzing the calculated radial distribution functions. The influence of counterions on the double helix structure has been studied in the fourth section. In the fifth section, the features of counterions structuring around the DNA macromolecule have been discussed.

## 2. Model and methods

A DNA fragment d(CGCGAATTCGCG), known as the Drew-Dickerson dodecamer [46], has been simulated. The atomic and schematic structures of the modelled DNA dodecamer are showed in Fig. 1.

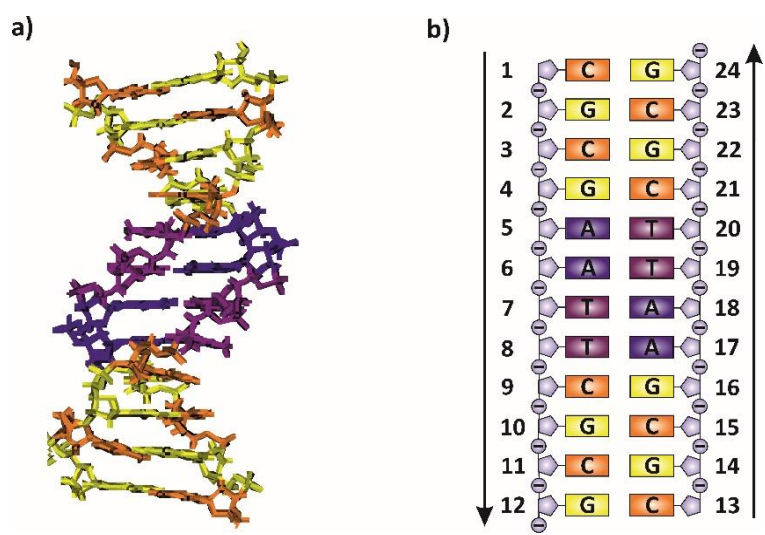

**Figure 1.** Modelled DNA fragment d(CGCGAATTCGCG). The colour corresponds to the type of nucleic base: adenine (A) -- dark blue, thymine (T) -- purple, guanine (G)-- yellow, cytosine (C) -- orange. a) Atomic structure of the fragment [46]. b) Schematic structure of the DNA fragment. The numeration of nucleotides is shown.

The counterions Na$^+$, K$^+$, and Cs$^+$ are added to the system. The ions are considered in two concentration regimes: i) the number of positively charged ions is equaled to the number of negatively charged phosphate groups of DNA backbone; ii) the chloride salt (NaCl, KCl, or CsCl) with a concentration of 0.5 M is added to the system of DNA with counterions neutralizing phosphate groups of the double helix. The oligomer is emersed into rectangular box (42Å x 42Å x 62Å) with about 2990 TIP3P water molecules. The periodic boundary conditions in all directions are used. Thus, three systems without added salt (Na0, K0, and Cs0) and three systems with added salt (Na05, K05, and Cs05) have been modelled. The parameters of these systems are showed in the Table 1.

**Table 1:** Parameters of the modelled systems. $Mt^+$ denotes the counterion type ($Na^+$, $K^+$, or $Cs^+$. $N_{Mt+}$ and $N_{Cl-}$ are the number of metal and chlorine ions in the system. $C$ is the concentration of added salt. $N_W$ is the number of water molecules.

| System name | $Mt^+$ | $N_{Mt+}$ | $N_{Cl-}$ | $C$ (M) | $N_W$ |
|---|---|---|---|---|---|
| Na0  | $Na^+$ | 22 | 0  | 0.0 | 2988 |
| Na05 | $Na^+$ | 50 | 28 | 0.5 | 2932 |
| K0   | $K^+$  | 22 | 0  | 0.0 | 2988 |
| K05  | $K^+$  | 50 | 28 | 0.5 | 2932 |
| Cs0  | $Cs^+$ | 20 | 0  | 0.0 | 2990 |
| Cs05 | $Cs^+$ | 50 | 28 | 0.5 | 2932 |

The computer simulations are performed within the framework of NAMD software package [47] and the CHARMM27 force field parameter set [48]. The integration time step is equaled to 2 fs. All bond length are fixed using SHAKE algorithm [49]. The long-range electrostatic interactions are treated using particle meshed Ewald method [50]. The switching and cutoff distances for the long-range interactions are taken equaled to 8 Å and 10 Å, respectively. The simulations are performed at the constant pressure (101325 Pa) and temperature 300 K. For pressure and temperature control Langevin dynamics is used for all heavy atoms (damping constant 5 $ps^{-1}$. The oscillation time and damping time constants for the Langevin piston are 100 fs and 50 fs, respectively.

The simulation protocol analogical to [29] is used. At the first stage of simulation, the minimization is made for the system with fixed heavy atoms (1000 steps). Then the system is minimized with fixed DNA atoms (2000 steps). At the second stage, the solvent is heated to a temperature of 300 K (15 ps) and equilibrated during 100 ps. At the third stage, the minimization is performed for the system with restrained DNA atoms by harmonic potential (2000 steps). Restrain coefficient is equaled to k=5 kcal/mol $Å^2$. Then the system is heated to a temperature of 300 K (30 ps) and equilibrated during 100 ps. After that, the atoms of DNA are set free and the system is equilibrated during 1 ns. The total duration of the MD trajectory is 12 ns. The first nanosecond is not taken into consideration in the further analysis of modelled systems.

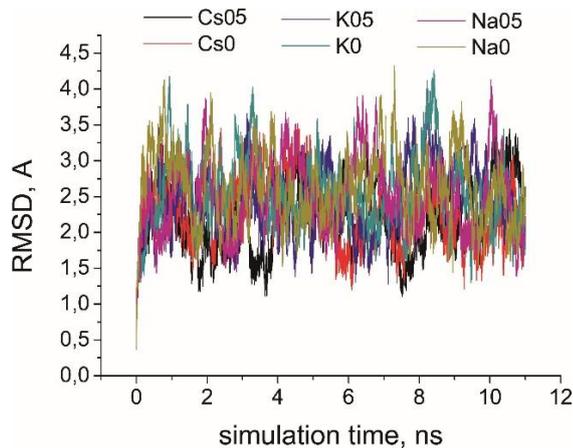

**Figure2.** Root mean square deviation (RMSD) of DNA atoms from the initial structure as a function on time. RMSD for different systems are shown by different colours.}

To perform the complete characterization of counterion dynamics around the DNA double helix the length of MD trajectory should be about several hundreds nanoseconds [21, 22,34]. However, for the comparative study of the dynamics of counterions of different type (in our case $Na^+$, $K^+$, and $Cs^+$, the length of MD trajectory may be shorter. Taking into consideration that the characteristic residence time of counterion in complex with the DNA atomic groups is usually about 1 ns [19-21], the MD trajectory of about 10 ns length may be appropriate. Thus, in the present study the lengths of MD trajectories for the modelled systems are taken equal to 12 ns.

To characterize the equilibrium state of the systems the deviation of DNA structure from the initial structure is analyzed using root mean square deviation (RMSD). The RMSD is defined as $\Delta R(t) \sqrt{\frac{1}{N}\sum_{\alpha=1}^{N}[\vec{r}_\alpha(t) - \vec{r}_\alpha(0)]^2}$ where $N$ is the number of considered atoms, $\vec{r}_\alpha(t)$ and $\vec{r}_\alpha(0)$ are the coordinates of atom $\alpha$ at the time $t$ and $t = 0$, respectively. RMSD of systems is derived using VMD program [51]. The obtained dependence of RMSD on time for all simulated systems is showed in Fig. 2. One can see that at the first nanosecond the RMSD values increase from 0 Å to about 2 Å and then fluctuate with respect to the average value. The amplitude of RMSD fluctuations is about 1 Å during all simulation trajectory. The obtained RMSD are appropriate for equilibrated systems at the nanosecond time scale.

## 3. Distribution of ions around DNA double helix

To analyze the distribution of counterions with respect to the DNA double helix the radial distribution functions (RDFs) of two type have been calculated using VMD program [51]. The first RDF type (phosphate RDF) characterizes the distribution of ions with respect to oxygen atoms O1 and O2 of the phosphate groups. The second RDF type (central RDF) characterizes the distribution of ions with respect to the atoms that are localized close to the center of *B*-DNA double helix (atoms N3 of purine and N1 of pyrimidine bases). The RDFs are calculated for all nucleotide pairs except that belonging to the ends of DNA fragment. To avoid the end effects the 1-21, 2-23, 12-13, and 11-14 nucleotide pairs (Fig. 1b) (ends of the dodecamer) are not considered. Using the obtained series of functions the averaged RDFs of each type have been defined. The averaged RDFs are showed in Fig. 3.

The phosphate RDFs describe the distribution of counterions outside the double helix. The obtained phosphate RDFs usually have two prominent and one weak maximums. The position of the first maximum is within the distance range (2.2 ÷ 3.1) Å (Table 2) which correlates with the ion size. The first peak characterizes the direct binding of ions to the oxygen atoms of the phosphate groups of DNA backbone. The second maximum at about (4.5 ÷ 7.3) Å characterizes the localization of ions in the grooves of the double helix and in the ion-hydrate shell of macromolecule. The third weak peak (at about 10 Å) characterizes the ions in the ion-hydrate shell of DNA.

**Table 2:** Positions of maximums of radial distribution functions (Å).

| System name | Na0 | Na05 | K0 | K05 | Cs0 | Cs05 |
|---|---|---|---|---|---|---|
| **Phosphate RDF** | | | | | | |
| I-peak | 2.2 | 2.2 | 2.6 | 2.6 | 3.1 | 3.1 |
| II-peak | 4.5 | 4.6 | 4.9 | 5.0 | 7.3 | 7.1 |
| **Central RDF** | | | | | | |
| I-peak | – | – | 4.9 | 4.9 | 5.2 | 5.3 |
| II-peak | 6.7 | 6.7 | 7.1 | 6.7 | 7.3 | 7.1 |

The central RDFs feature the localization of counterions with respect to the double helix axis. The obtained central RDFs may have one or two peaks. The first peak is at about the distance 5 Å, while the second peak is at about 7 Å from the helix center. These peaks characterize the localization of counterions in the grooves of the double helix. The first peak arises due to the ions contacting with the atoms of bases directly at the bottom of the grooves. The second peak characterizes the localization of ions at the top edge of the grooves.

Our results have showed that the distribution of Na$^+$ counterions in the both Na0 and Na05 systems is characterized by the first intensive and the second small maximums in the phosphate RDF (Fig. 3a). The positions of these peaks are at about 2.2 Å and 4.5 Å from the phosphate group, respectively (Table 2). The central RDF has a band with low intensity at about 7 Å from the double helix center. The first intensive peak in the phosphate RDF and the band in the central RDF indicate that sodium counterions interact with the phosphate groups and do not enter the grooves deeply. The Na$^+$ ions with water molecules of its hydration shell may be localized at top edge of the grooves.

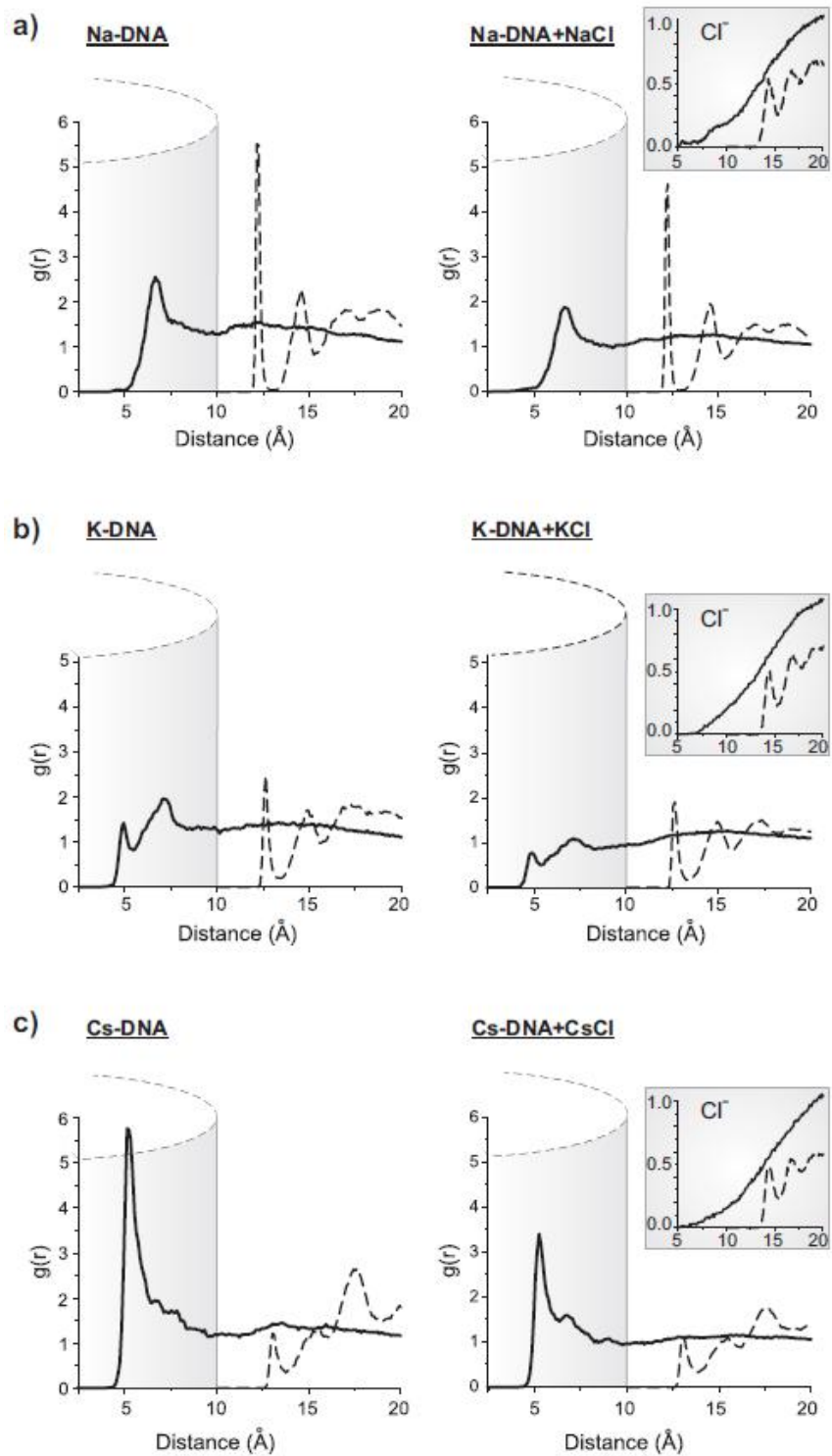

**Figure 3.** Averaged radial distribution functions (RDFs). Central RDFs (solid lines) and phosphate RDFs (dashed lines) describe the distribution of the ions with respect to the double helix axis and backbone phosphates, respectively. The phosphate RDFs are shifted for 10 Å to illustrate the characteristic radius of *B*-DNA. The cylinder schematically shows the surface of the double helix. a) The RDFs of $Na^+$ counterions in Na0 and Na05 systems. b) The RDFs of $K^+$ counterions in K0 and K05 systems. c) The RDFs of $Cs^+$ counterions in Cs0 and Cs05 systems. In insets the RDFs for $Cl^-$ ions are shown.

In the case of K0 and K05 systems, the maximums of phosphate RDFs are not as intensive as in the systems with sodium ions (Fig. 3b). At the same time, the central RDF has two peaks at the distances about 5 Å and 7 Å (Table 2) characterizing a direct interaction of the ions with atoms of bases and localization of ion at the top edge of grooves, respectively. The interaction of $K^+$ with the phosphate groups is not dominant.

The radial distribution functions of Cs0 and Cs05 systems are characterized by an intensive peak in the central RDF at the distance about 5 Å and by a shoulder at about 7 Å (Fig. 3c). The phosphate RDFs are characterized by low first and second peaks and a rather strong band at about 7 Å (Table 2). Such structure of RDFs indicate that $Cs^+$ ions are localized mostly in the grooves of the double helix and interact directly with the atoms of bases.

The distribution of $Cl^-$ ions in the Na05, K05, and Cs05 systems is rather similar. The central RDFs have no maximums and gradually increase as the distance from the double helix center grows. The phosphate RDFs are characterized by a low maximum at a distance of about 5 Å that is more than twice greater than the ion radius (insets in Fig. 3). Thus chlorine anions tend to be localized at large distances from the DNA double helix.

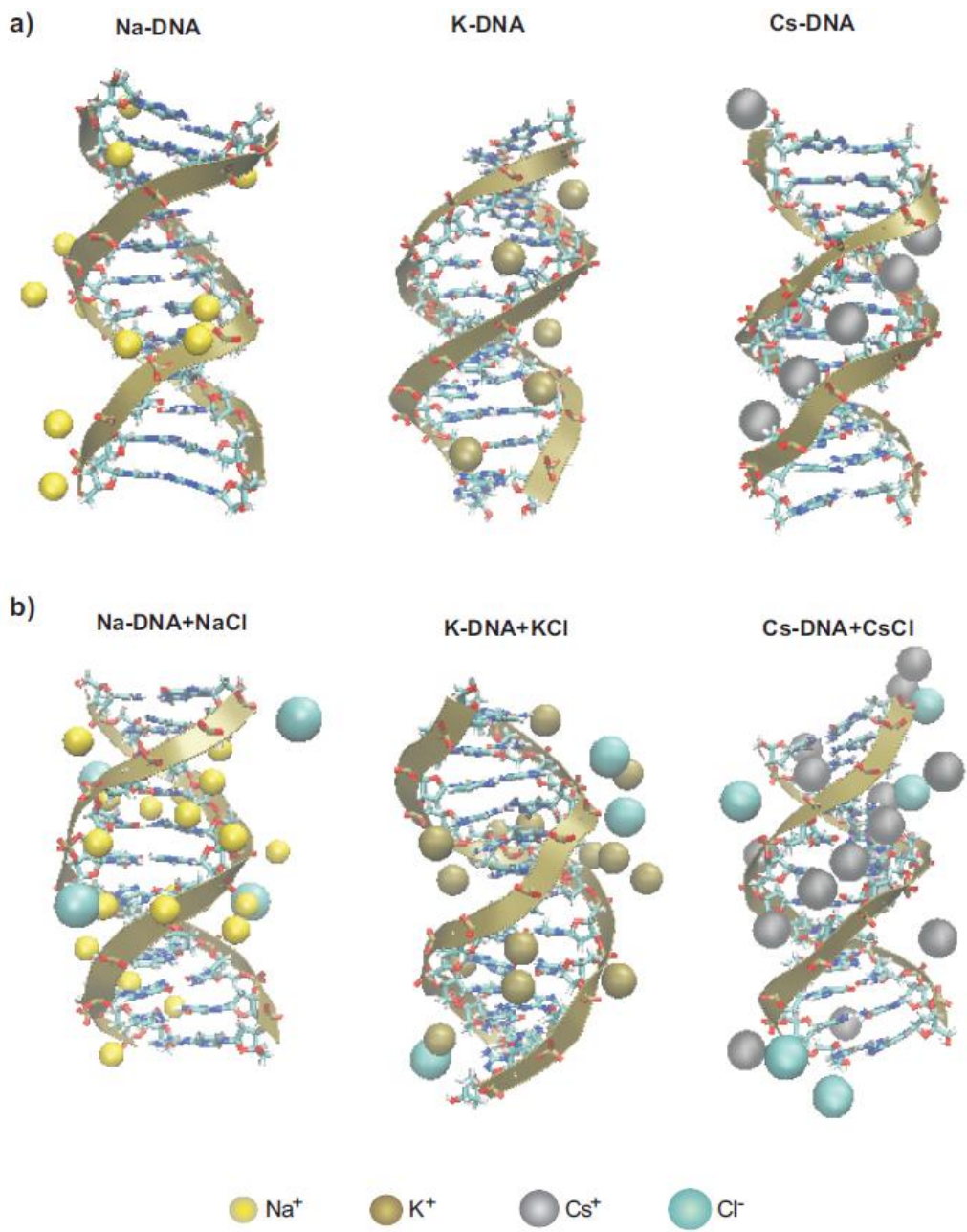

**Figure 4.** Snapshots of DNA double helix with counterions. a) DNA with $Na^+$, $K^+$, and $Cs^+$ counterions in Na0, K0, and Cs0 systems, respectively. b) DNA with added NaCl, KCl, and CsCl salts in Na05, K05, and Cs05 systems.

The analysis of the radial distribution functions shows that the counterions usually interact with the atoms O1 and O2 of phosphates, the atoms N3, N7, and O2 of nucleic bases, and the atom O4 of deoxyribose. The snapshots illustrating the localization of $Na^+$, $K^+$, and $Cs^+$ counterions on DNA double helix are showed on Fig. 4. A visual examination has showed that in the case of salt free system, the sodium counterions are localized near the phosphate groups from the outside of the double helix and in the grooves of the macromolecule (Fig. 4a). In the minor groove of the double helix, the $Na^+$ ions can be localized between phosphate groups of different strands at the top edge of the groove. The interaction of sodium ions with the atoms of nucleic bases occurs via water molecules. The potassium ions in the K0 system can interact mostly to the phosphate groups from the grooves side of macromolecule. In contrast to the $Na^+$ ions, the $K^+$ ions an also interact directly with the atoms of nucleic bases. The cesium counterions in the Cs0 system can penetrate deeply into the grooves. In the minor groove, they interact with N3 and O2 atoms of purine and pyrimidine bases. From time to time, the regular structure of $Cs^+$ counterions are observed in the minor groove of the double helix. The distance between structured $Cs^+$ ions in the minor groove of the double helix is about 7 Å.

In the systems with added $Mt^+Cl^-$ salt (Na05, K05, and Cs05), the interaction sites of counterions with DNA double helix are the same as in salt free solutions. The chlorine ions form the associates with the metal ions. The cation-anions pairs can approach rather close to macromolecule and even be incorporated into the double helix (Fig.4b). The interaction of cations with the central base pairs of DNA fragment is more intensive, while the anions more often come to the ends of the double helix.

The analysis of obtained results shows that derived RDFs have the shape similar to that obtained in MD studies with much longer simulation trajectory [21, 22, 28, 30, 34], but the intensities of RDFs maximums are slightly different. The detected DNA-counterion contacts and their lifetimes correlate with the values obtained earlier [21, 28-30]. Thus, the conducted MD research describes well (at least, in terms of qualitative aspects) the features of counterion structuring around DNA double helix.

# 4. Transformations of the double helix structure under the influence of counterions

The counterions interacting with DNA atoms induce the conformational changes of the double helix. To analyze the influence of counterions on conformational state of the macromolecule, the structure parameters of the double helix have been calculated. The DNA conformational parameters have been derived with the help of Curves+ program [52]. Using the obtained data the averaged on time parameters of the double helix grooves have been calculated. The dependence of groove widths and depths on the base pair level are showed in Figs. 5 and 6.

The obtained dependence of groove parameters on base pair level and the influence of counterions correlate with the results of early molecular dynamics simulations [27, 29-32, 34]. In the case of salt free systems, the width of the major and minor grooves changes within the range (8 ÷ 12) Å and (5.5 ÷ 11) Å, respectively (Fig. 5a, 6b). The minimal values of groove width correspond to the central base pairs, while the largest values are characteristic for the ends of the DNA fragment. For the central base pairs the values of groove width are the lowest in the case of $Cs^+$ ions. In the case of $Na^+$ and $K^+$ ions width of the grooves is almost similar.

The width of the grooves in the case of systems with added salt varies within the range (6 ÷ 13) Å and has the same character of dependence on base pair level as that in the salt free systems (Figs. 6a, 6b). The groove depth is approximately constant in the central part of the dodecamer (about 6 Å) and changes in its ends for all systems (Figs. 5c, 5d and Figs. 6c, 6d).

To characterize the macromolecule elasticity the time averaged angle of the double helix bending has been calculated for DNA in the modelled systems (Table 3). One can see that the average angle values are rather large and vary within the range (10 ÷ 21)°. The bending angle is especially large in the case of K05 and Cs0 systems, while in the case of Na0 and K0 systems it has the lowest value. The bending angle is related with the persistence length of the macromolecules. Increasing the double helix bending angle should induce a decrease in the persistence length of DNA. The dependence of persistence length on counterion type and concentration has been observed experimentally [4].

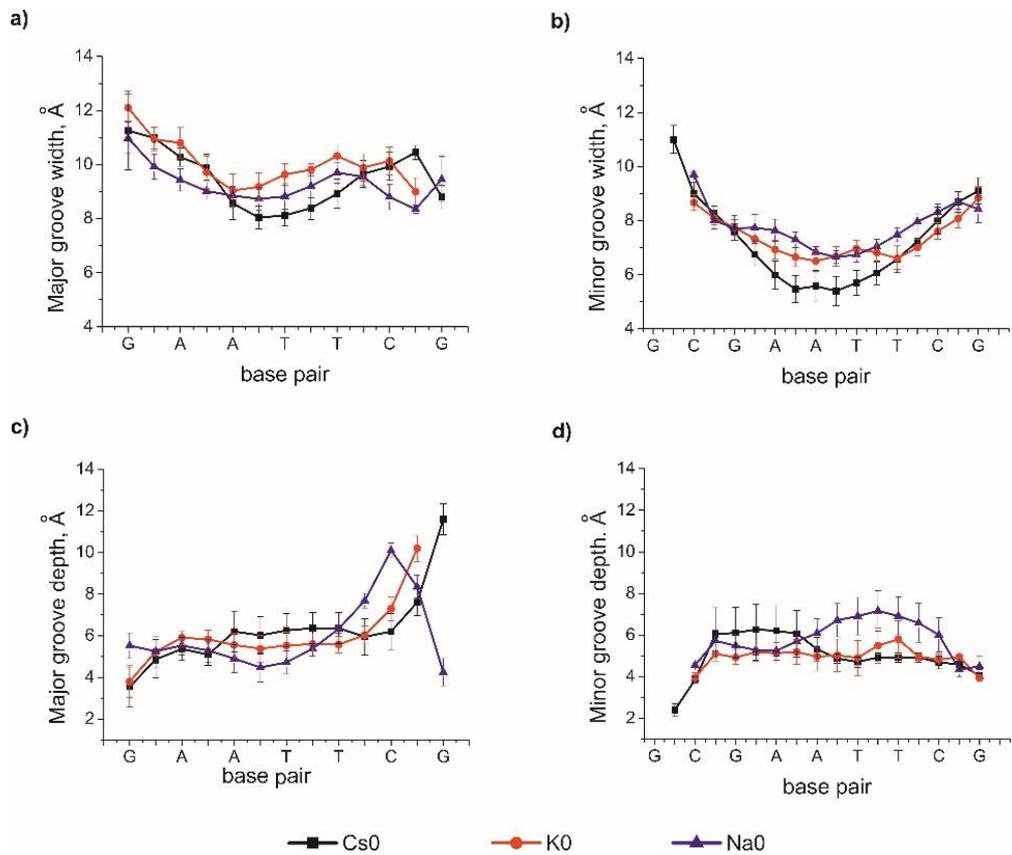

**Figure5.** Average grooves parameters for Na0, K0, and Cs0 systems. a) Major groove width. b) Minor groove width. c) Major groove depth. d) Minor groove depth.

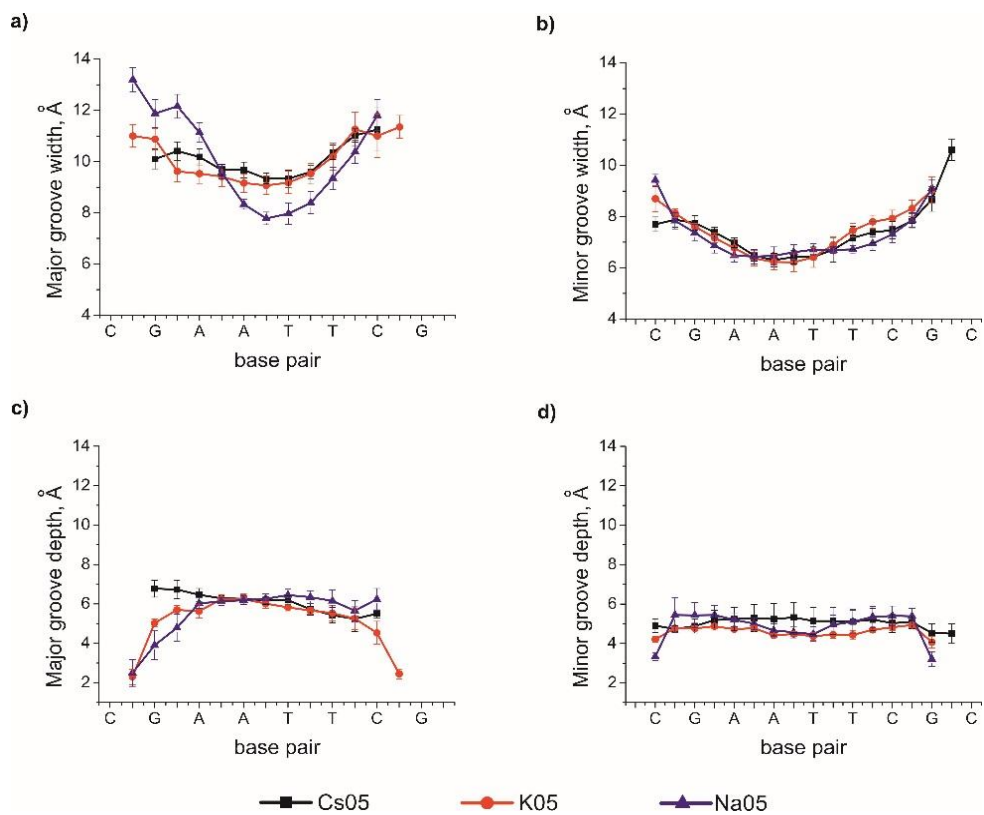

**Figure6.** Average grooves parameters for Na05, K05, and Cs05 systems. a) Major groove width. b) Minor groove width. c) Major groove depth. d) Minor groove depth.

Table 3: Values of total bend of DNA macromolecules for different systems.

| System | Na0 | Na05 | K0 | K05 | Cs0 | Cs05 |
|---|---|---|---|---|---|---|
| Angle (°) | 11 ± 4 | 15 ± 7 | 11 ± 7 | 21 ± 9 | 21 ± 5 | 12 ± 7 |

Our analysis of DNA structure parameters has showed that the conformational changes of the double helix are the most prominent in the center of the modelled DNA dodecamer where the localization of counterions is the most often. At the same time, the conformational changes may occur due to the length effects of the polynucleotide and due to the intrinsic properties of the AATT sequence of DNA that is in the central part of Drew-Dickerson dodecamer (Fig.1). All these factors of the double helix structural transformations may be interrelated.

## 5. Discussion

The results of conducted MD study have showed that $Na^+$, $K^+$, and $Cs^+$ counterions interact with DNA in different manner and the relative position of ions on double helix is essentially different. Sodium ions are bound mostly to the phosphate groups of the DNA backbone and may enter the grooves of the double helix with water molecules of the hydration shell. The potassium ions are localized mostly in the grooves of the double helix. The interaction of $K^+$ with the phosphates is not so essential as in the case of $Na^+$. The cesium ions penetrate deeply into the grooves of the double helix and interact directly to the atoms of bases.

The reason for such different character of interaction of alkali metal ions with DNA is considered due to different size of ions and their different solvation type [19]. The radiuses of alkali metal ions are known to increase in the following order: $Li^+<Na^+<K^+<Rb^+<Cs^+$, while their hydration radiuses have opposite gradation [53]. As the result, energy of hydration is positive in case of $Li^+$ and $Na^+$ ions, and is negative in case of $K^+$, $Rb^+$ and $Cs^+$ ions [54]. That is why, the water molecules of hydration shell of the sodium and lithium ions are constrained that does not allow the ions to enter the grooves of the double helix deeply and to interact directly with the atoms of bases. The smallest alkali metal ion $Li^+$ interacts with the oxygen atoms of DNA phosphates and form the common hydration shell which provides a long living complex with the double helix [19]. In contrast, the water molecules around the cesium ions are not so constrained by ions, and $Cs^+$ can be easily dehydrated and pass to the bottom of DNA grooves. The properties of the rubidium ions in DNA solutions have been showed to be rather similar to the $Cs^+$ ions [19, 34, 35].

Due to the electrostatic repulsion, the negatively charged chlorine ions tend to be localized at large distances from DNA polyanion. However, some $Cl^-$ ions has been detected close to DNA surface and even inside the grooves of the macromolecule. Intrusion of the $Cl^-$ into the double helix occurs due to the formation of neutral cation-anion pairs with the counterions already condenses on DNA. A recent MD study of ionic aggregates in DNA solutions has showed that the formation of large ionic clusters is characteristic for different force fields, but not for the CHARMM force field used in the present work [55]. Thus the formation of cation-anion pairs in our MD simulations is not an artifact of the method.

The structure of the DNA double helix is not uniform for all base pairs. The grooves of the double helix shrink at the center of the macromolecule fragment and the double helix is bent as a whole. The observed conformational transformations of the macromolecule occur due to both intrinsic properties of the AATT sequence of the DNA fragment and the influence of the counterions interacting with the double helix. The length effects of DNA fragment may be also essential.

The relative disposition of $Na^+$ and $K^+$ ions on DNA macromolecules change dynamically. The lifetime of these ions tethered to the phosphate groups of the double helix is equal or less then 0.5 ns. The cesium ions can spend rather long time in complex with DNA atomic groups (about several nanoseconds) and, from time to time, form a regular structure in the minor groove of the double helix. The characteristic distance between structured $Cs^+$ ions is about 7 Å. The structure of $Cs^+$ counterions with DNA may be considered as the ionic lattice. The existence of such lattice structure for DNA with alkali metal ions was predicted in the previous works [39-43].

## 6. Conclusions

In the present research, the dynamical and structural features of DNA-counterion system have been studied by the molecular dynamics method. The MD simulations have been made for different systems consisting of the molecular fragment d(CGCGAATTCGCG) and counterions of some type ($Na^+$, $K^+$, or $Cs^+$). The systems have been considered in the regimes without excess salt and with the different salts (0.5 M of NaCl, KCl or CsCl) added. The results have showed that the $Na^+$ ions interact mostly to phosphate groups of the DNA backbone. The $K^+$ ions are localized mostly in grooves of the double helix, while the $Cs^+$ ions penetrate deeply inside the grooves of the double helix and interact directly to the atoms of bases. The difference in counterion interactions with DNA is considered as a consequence of different ion hydration. The addition of extra salt to the system does not change the character of interaction of counterions with DNA. The chlorine anions are usually localized at large distances from DNA polyanion, but some $Cl^-$ ions have been detected near atomic groups of the double helix that is due to the formation of neutral pairs with the counterions already condensed on DNA. The interactions of counterions with DNA induce local changes in the double helix structure. The counterions dynamically change their positions on DNA macromolecule. The lifetime of $Na^+$ and $K^+$ ions near the phosphate groups of the double helix is equal to or less than 0.5 ns. In contrast, the lifetime of cesium ions in DNA minor groove may reach several nanoseconds. In this time scale, the regular structure of $Cs^+$ in the DNA double helix is formed, which may be considered as ionic lattice.

*The research is supported by the National Academy of Sciences of Ukraine (project 0114U000687).*